\documentclass{article}[12pt]
\usepackage{amscd,amsmath,amsthm,amsfonts,amssymb}
\usepackage{graphicx}
\usepackage{multicol}
\usepackage{color}

\def\sech{{\rm sech}}
\def\cosh{{\rm cosh}}
\def\tanh{{\rm tanh}}

\def\ri{{\rm i}}

\def\re{{\rm e}}
\def\PT{$\mathcal{PT}$}
\def\P{\mathcal{P}}

\def\[{\begin{equation}}
\def\]{\end{equation}}

\begin{document}
\title{\bf Construction of non-\PT-symmetric complex potentials with all-real spectra}
\author{Jianke Yang\footnote{Email: jyang@math.uvm.edu}  \\ Department of Mathematics and Statistics \\ University of Vermont \\ Burlington, Vermont, USA}
\date{ }
\maketitle

\noindent
\textbf{Abstract:}

\vspace{0.2cm}
We review recent work on the generalization of \PT symmetry. We show that, in addition to \PT-symmetric complex potentials, there are also large classes of non-\PT-symmetric complex potentials which also feature all-real spectra. In addition, some classes of these non-\PT-symmetric potentials allow phase transitions which do or do not go through exceptional points. These non-\PT-symmetric potentials are constructed by a variety of methods, such as the symmetry and supersymmetry methods and the soliton theory. A generalization of \PT symmetry in multi-dimensions is also reviewed.

\section{Introduction}
Linear paraxial propagation of light in an optical waveguide is governed by the Schr\"{o}dinger equation \cite{Kivshar_book,Musslimani2008,Yang_book}
\begin{equation} \label{e:SE}
\ri \Psi_z + \Psi_{xx} + V(x) \Psi=0,
\end{equation}
where $z$ is the distance of propagation, $x$ is the transverse coordinate, $V(x)$
is a complex potential whose real part is the index of refraction and the imaginary part represents gain and loss
in the waveguide. This same equation also arises in non-Hermitian quantum mechanics \cite{Bender1998,Ali}
and Bose-Einstein condensates \cite{BECbook}, in which case $z$ is the time variable. Looking for eigenmodes of the form $\Psi(x,z) = \re^{\ri \mu z} \psi(x)$ we arrive at the eigenvalue problem
\begin{equation} \label{Lpsi}
L\psi=\mu \psi,
\end{equation}
where $L = \partial_{xx} + V(x)$ is a Schr\"odinger operator, and $\mu$ is an eigenvalue.

All-real spectrum for this Schr\"odinger operator with a complex potential is a sought-after property not only in non-Hermitian quantum mechanics, but also in optics and Bose-Einstein condensates. In quantum mechanics, $\mu$ is the energy level, which must be real in order for the system to be physically meaningful. In optics and Bose-Einstein condensates, an all-real spectrum of a complex potential is significant because it allows a dissipative optical system with gain and loss to behave like a conservative system.

Bender and Boettcher \cite{Bender1998} first observed that, if the Schr\"odinger operator $L$ is parity-time (\PT) symmetric, i.e., it is invariant under the joint transformations of $x \to -x$ and complex conjugation, then its spectrum can be all-real. This \PT symmetry is equivalent to the condition on the complex potential that
\[
V^*(x)=V(-x),
\]
where the asterisk * represents complex conjugation. In optics, this condition means that the refractive index needs to be an even function in space, and the gain-loss profile needs to be an odd function in space. A simple reason for this all-real spectrum of a \PT-symmetric potential is that its eigenvalues always appear in complex conjugate pairs. This is because for such potentials, if $\mu$ is an eigenvalue with eigenfunction $\psi(x)$, then by taking the complex conjugate of Eq. (\ref{Lpsi}) and switching $x\to -x$, we see that $\mu^*$ would also be an eigenvalue with eigenfunction $\psi^*(-x)$. This eigenvalue symmetry restricts the appearance of complex eigenvalues and facilitates the realization of an all-real spectrum. But this \PT symmetry does not necessarily guarantee an all-real spectrum, and phase transition can occur when conjugate pairs of complex eigenvalues appear in the spectrum \cite{Musslimani2008,Bender1998,Ahmed2001}.

\PT symmetry has found many optical applications, such as unidirectional reflectionless metamaterials \cite{Feng2013}, \PT lasers \cite{PTlaser_Zhang,PTlaser_CREOL}, and non-reciprocity in \PT-symmetric whispering-gallery microcavities \cite{Peng}. In these optical applications, the refractive index and gain-loss profiles of the waveguide were carefully designed so as to respect \PT symmetry. In a \PT setting, the gain-loss profile must be anti-symmetric, which could be restrictive. The pursuit of non-\PT-symmetric potentials with more flexible gain-loss profiles and all-real spectra is thus an interesting question.  In recent years, various techniques have been developed to construct non-\PT-symmetric potentials with all-real spectra, and they will be reviewed in this article (a brief review on some of these results could also be found in \cite{Yang_review}).

\section{Non-\PT-symmetric potentials with all-real spectra and exceptional-point-mediated phase transition} \label{sec:typeI_II}

To derive non-\PT-symmetric complex potentials with all-real spectra, one strategy is to impose an operator symmetry in order to guarantee conjugate-pair eigenvalue symmetry \cite{Yang2016}. Like the case of \PT-symmetric potentials, this conjugate-pair eigenvalue symmetry guarantees that either the spectrum of $L$ is all-real, or a phase transition occurs when pairs of complex eigenvalues appear.

To execute this strategy, we observe that if there exists an operator $\eta$ such that $L$ and its complex conjugate $L^*$ are related by a similarity relation
\begin{equation} \label{Eq:LSimilar}
\eta L =  L^* \eta,
\end{equation}
then the eigenvalues of $L$ would come in conjugate pairs if the kernel of $\eta$ is empty \cite{Yang2016}. This operator relation resembles the condition for pseudo-Hermiticity \cite{Ali}, but we do not require $\eta$ to be invertible here.

If we let $\eta = \mathcal{P}$, where $\mathcal{P}$ is the parity operator $x \mapsto -x$, then this $\eta$ operator has an empty
kernel, and the similarity condition (\ref{Eq:LSimilar}) yields
$V(-x)  =V^*(x)$, which recovers the well-known class of
\PT-symmetric potentials. However, when branching out to different
choices of $\eta$, a completely real spectrum can be obtained for an
arbitrary choice of the gain-loss distribution by a judicious
construction of the index of refraction. This will be demonstrated
below where $\eta$ is chosen as a differential operator.

\subsection{Type-I potentials}
First, we consider the simplest choice of a differential $\eta$ operator,
\begin{equation} \label{e:eta1}
\eta = \partial_x + a(x).
\end{equation}
Substituting this $\eta$ and operator $L$ into the
similarity condition (\ref{Eq:LSimilar}), we get the following two
equations
\begin{equation} \label{e:iG}
a_x=\ri \hspace{0.06cm}\mbox{Im}(V), \qquad a_{xx}-V_x=(a^2)_x.
\end{equation}
The second equation can be integrated once, and we get
\[
a_{x}-V=a^2+\xi_0,
\]
where $\xi_0$ is a constant. Utilizing (\ref{e:iG}), this equation becomes
\begin{equation} \label{e:n}
\mbox{Re}(V)=-a^2-\xi_0.
\end{equation}
Eqs. (\ref{e:iG})-(\ref{e:n}) show that $a(x)$ is a purely imaginary function, and $\xi_0$ is a
real constant. Denoting $a(x)=\ri g(x)$, where $g(x)$ is an arbitrary
real function, we get $\mbox{Re}(V)=g^2(x)-\xi_0$ and $\mbox{Im}(V)=g'(x)$, with the prime representing the derivative. The constant
$\xi_0$ can be eliminated by a gauge transformation to Eq.~(\ref{e:SE}), and thus the resulting complex potential is
\begin{equation} \label{e:V1}
V(x)=g^2(x)+\ri g'(x).
\end{equation}
These potentials were called type-I potentials in \cite{Yang2016}. They generalized the potentials of the same form in \cite{Tsoy2014,Wadati2008}, where special choices of the $g(x)$ function were taken (see sections \ref{sec_typeI_II_all_real} and \ref{sec:soliton} for reviews).

Compared to \PT-symmetric potentials, a distinctive feature of these type-I potentials is that the gain-loss profile $g'(x)$ is now arbitrary since $g(x)$ is arbitrary. But due to the symmetry relation (\ref{Eq:LSimilar}), the spectra of these potentials can still be all-real, just like \PT-symmetric potentials. This possibility for all-real spectra for arbitrary gain-loss profiles is made possible by a judicious choice of refractive indices in relation to the gain-loss profiles.

\begin{figure}[htb]
\begin{center}
\includegraphics[width=0.8\textwidth]{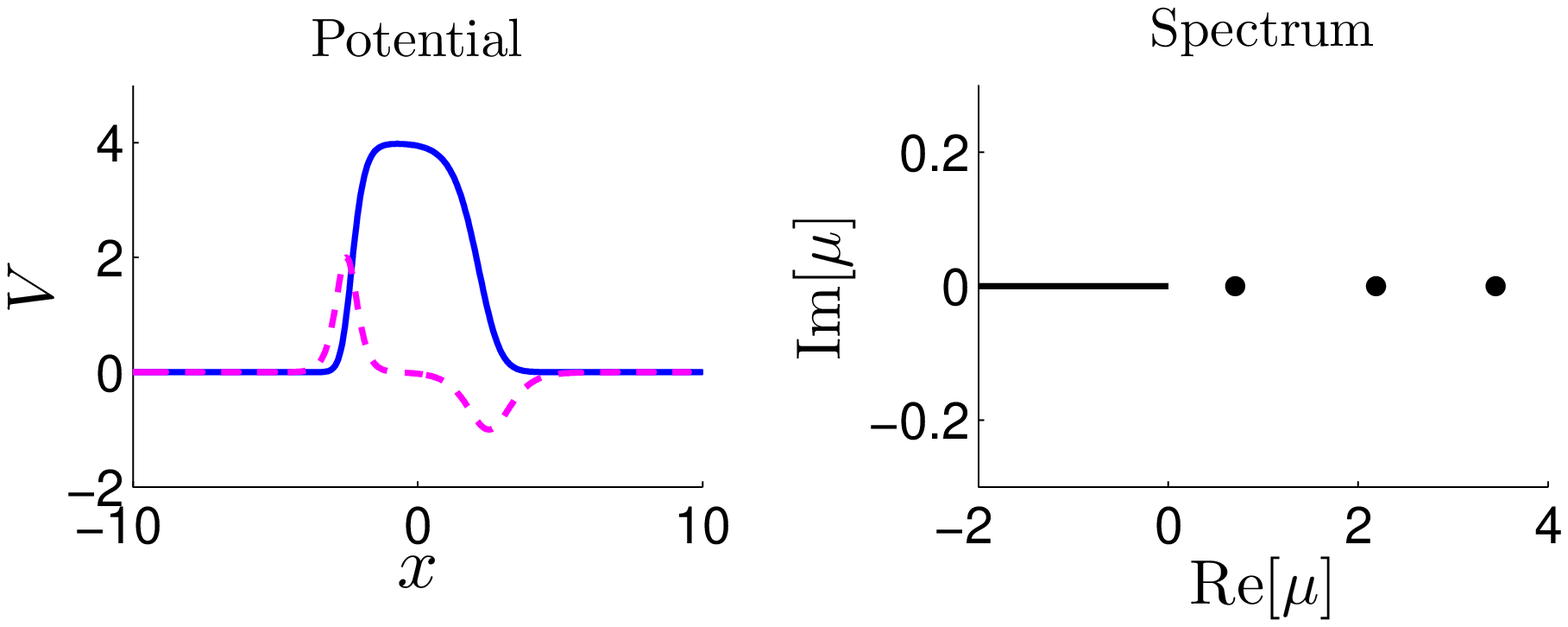}

\vspace{0.1cm}
\includegraphics[width=0.8\textwidth]{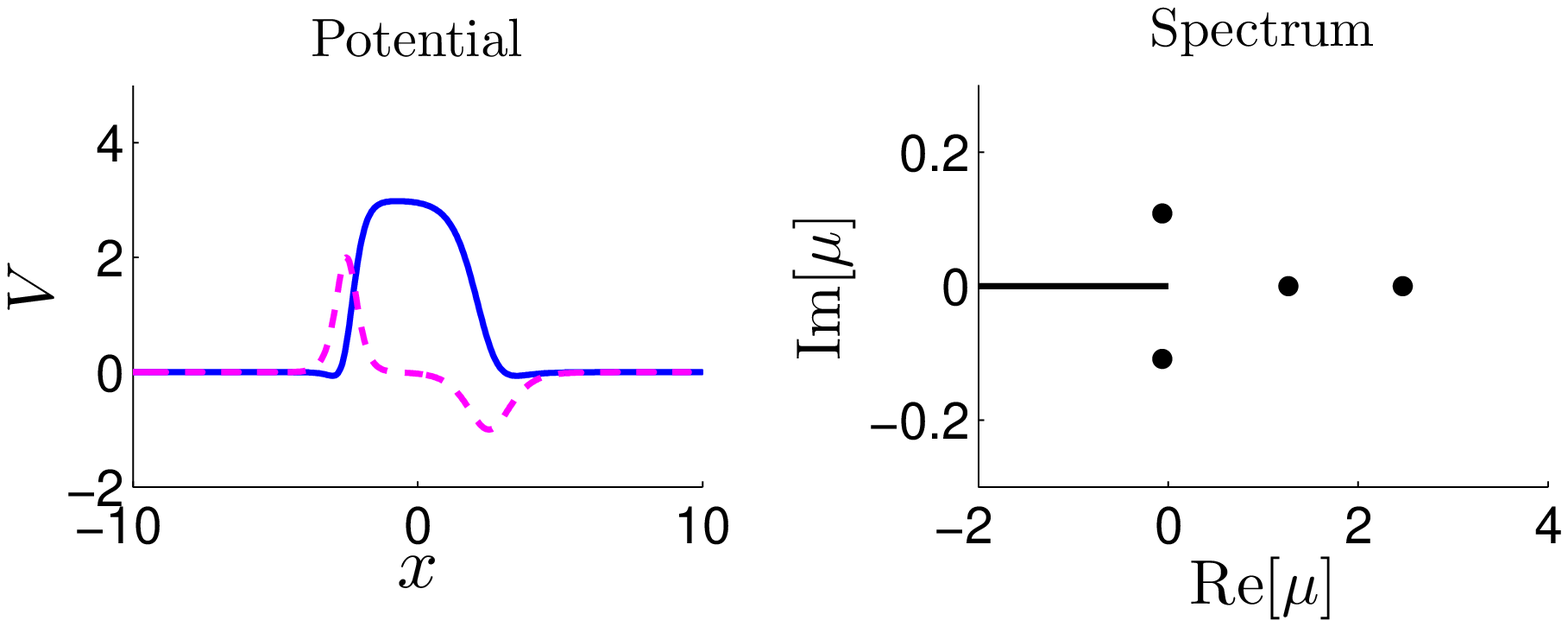}
\end{center}
\caption{Spectra of type-I potentials
(\ref{e:V1}) with $g(x)$ given in (\ref{Eq:g}). Upper row: $c_0 = 0$; lower row: $c_0 = -0.3$. In the potentials,
the solid blue line is Re($V$), and the dashed pink line is Im($V$). Adapted from \cite{Yang2016}. }
\label{Fig:FirstOrder}
\end{figure}

As an example, we take
\begin{equation} \label{Eq:g}
g(x)=\tanh [2(x+2.5)]-\tanh (x-2.5)+c_0,
\end{equation}
where $c_0$ is a real constant. In Fig. \ref{Fig:FirstOrder}, we show two
potentials of the form \eqref{e:V1}, with the $c_0$ value taken as 0 and $-0.3$ respectively.  In the upper row, the
potential with $c_0=0$ has a completely real spectrum, and
increasing $c_0$ will maintain the reality of the spectrum as more
discrete eigenvalues bifurcate off the edge of the continuous
spectrum. However, as $c_0$ is decreased, the spectrum will undergo a phase
transition at $c_0\approx -0.181$, where a pair of real eigenvalues and their eigenfunctions coalesce and form an exceptional point at $\mu\approx 0.056$. This exceptional point then
bifurcates off the real axis and creates a pair of complex eigenvalues afterwards. This can be seen in the lower row of Fig. \ref{Fig:FirstOrder} with $c_0=-0.3$. We stress that this phase transition is induced by going through an exceptional point, which is a common scenario for phase transition \cite{Bender1998,Ahmed2001}.

\small
\normalsize

\subsection{Type-II potentials}  \label{sec_typeII}
Type-I potentials (\ref{e:V1}) come from taking the
simplest form of a differential $\eta$ operator [i.e., a first-order
operator (\ref{e:eta1})]. By increasing the order of this
differential operator, more families of potentials arise. Let $\eta$
now be a second-order operator,
\begin{equation}
\eta = \partial_{xx} + a(x) \hspace{0.04cm} \partial_{x} + b(x).
\end{equation}
Inserting this $\eta$ into the similarity condition
(\ref{Eq:LSimilar}) and collecting coefficients of the same order of
derivatives on the two sides of this condition, we get
\begin{center}
\begin{tabular}{|c||c|c|}
\hline
&$\eta L$ &$L^* \eta$ \\
\hline
$\partial_x^4$ &$ 1$ & $1$\\
$\partial_x^3$ & $a$ & $a$  \\
$\partial_x^2$ & $ V $ & $V^*+ 2a'$\\
$\partial_x^1$ &$V a+2V'$&$V^* a+ a''+ 2b'$\\
$\partial_x^0$ &$Vb+ V' a+V''$&$V^*b+ b''$\\
\hline
\end{tabular}
\end{center}

\vspace{0.3cm}
\noindent
Setting these coefficients in $\eta L$ and $L^*\eta$ to match each
other, we get a system of equations which can be solved from top to
bottom. From the $\partial_x^2$ coefficients, we get $a'(x)=\ri\hspace{0.06cm}\mbox{Im}(V)$.
Setting $a(x) = \ri g(x)$, where $g(x)$ is a real function, we
obtain $\mbox{Im}(V)=g'(x)$. Inserting this $a(x)$ formula into the
$\partial_x^1$ equation and integrating once, we get
\[
b = \mbox{Re}(V) -\frac{1}{2} g^2 + \frac{\ri}{2}  g' + c_1,  \nonumber
\]
where $c_1$ is a constant.

Now we insert these $a(x)$ and $b(x)$ solutions into the $\partial_x^0$ equation. After simple algebra, this equation becomes
\[
[\mbox{Re}(V)g^2]' = g^3g' -\frac{1}{2} g''' g- 2 c_1 g'g,   \nonumber
\]
from which we can solve the refractive index $\mbox{Re}(V)$ as
\begin{equation}  \label{e:nII}
\mbox{Re}(V) = \frac{1}{4} g^2+ \frac{g'^2-2 g''g +c_2}{4g^2} - c_1,   \nonumber
\end{equation}
where $c_2$ is a real constant. The overall constant $c_1$ can be removed without loss of generality.

Putting the above results together, we get potentials
\begin{equation}  \label{Eq:PotentialTypeII}
V(x) = \frac{1}{4}g^2+ \frac{g'^2-2g''g +c_2}{4g^2} +\ri g',
\end{equation}
where $g(x)$ is an arbitrary real function, and $c_2$ is an arbitrary real constant.
These potentials were called type-II potentials in \cite{Yang2016}, and they generalized potentials of the same form but with $c_2\le 0$ in \cite{Andrianov} (see sections \ref{sec_typeI_II_all_real} and \ref{sec_SUSY} for reviews).

Like type-I potentials, the gain-loss profile $g'(x)$ of these type-II potentials is also arbitrary, but their spectra can still be all-real due to the symmetry relation~(\ref{Eq:LSimilar}).

\begin{figure}[htb]
\begin{center}
\includegraphics[width=0.8\textwidth]{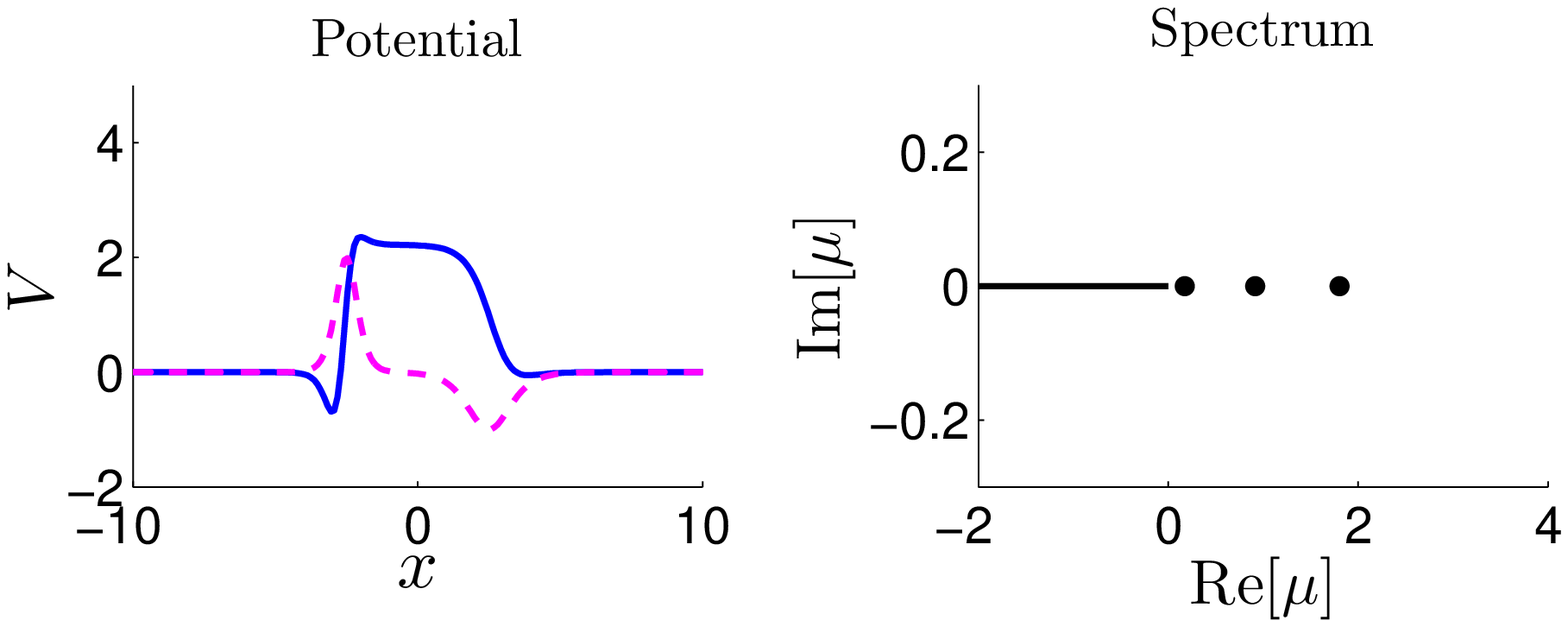}

\vspace{0.1cm}
\includegraphics[width=0.8\textwidth]{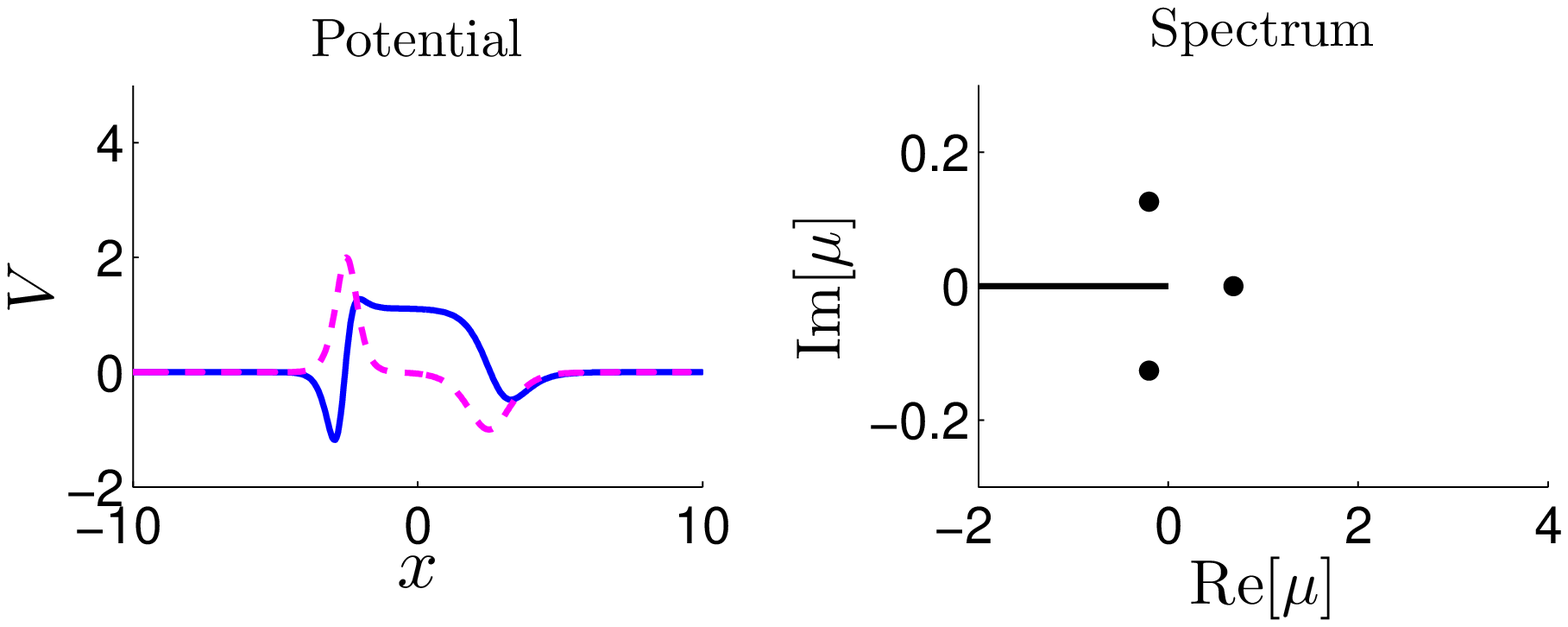}
\end{center}
\caption{Spectra of type-II potentials (\ref{Eq:PotentialTypeII}) with $g(x)$ given in (\ref{Eq:g}).
Upper row: $c_0 = 1$ and $c_2=-1$; lower row: $c_0 = 1$ and
$c_2=4$. Adapted from \cite{Yang2016}.} \label{Fig:SecondOrder}
\end{figure}

As an example, we take the same function $g(x)$ as in (\ref{Eq:g}). The spectrum of this potential
with $c_0=1$ and $c_2=-1$ is completely real, see
Fig.~\ref{Fig:SecondOrder} (upper row). Fixing $c_0$ and decreasing
$c_2$ will maintain the all-real spectrum. If $c_2$ is increased
above a certain threshold (which is approximately 2.535), a phase transition occurs, where
a pair of real eigenvalues coalesce and form an exceptional point, which then
bifurcates off the real axis and creates a pair of complex eigenvalues afterwards. This can be seen in the lower row of
Fig.~\ref{Fig:SecondOrder} for $c_2=4$. In this lower
row, an overall real constant $(c_0^2+c_2/c_0^2)/4$ has been subtracted
from the potential (\ref{Eq:PotentialTypeII}) so that the refractive
index drops to zero at infinity. Again, the phase transition here is induced by going through an exceptional point.

\small
\normalsize

\subsection{Potentials with strictly all-real spectra} \label{sec_typeI_II_all_real}

For the two types of potentials (\ref{e:V1}) and (\ref{Eq:PotentialTypeII}), an exceptional-point-mediated phase transition is in general possible (see Figs. \ref{Fig:FirstOrder} and \ref{Fig:SecondOrder}). But under certain restrictions on these potentials, all-real spectra can be guaranteed.

For type-I potentials (\ref{e:V1}), it was shown by Tsoy \emph{et al.} \cite{Tsoy2014}
that, if $g(x)$ is a single-humped localized real function, then its spectrum is strictly real. This result was based on an observation by Wadati \cite{Wadati2008} that the Zakharov-Shabat (ZS) spectral problem \cite{ZS71}
\begin{equation}
\label{ZS0}
v_{1x}+\ri \zeta v_1=g(x)v_2, \,\,\, v_{2x}-\ri \zeta v_2=-g(x)v_1,
\end{equation}
with $\zeta$ being a spectral parameter, can be transformed into the Schr\"odinger eigenvalue problems
\[
\psi_{xx}+V(x,t)\psi=\mu\psi,  \quad
\phi_{xx}+V^*(x,t)\phi=\mu \phi,         \nonumber
\]
with $V(x)$ being the type-I potential (\ref{e:V1}) and $\mu=-\zeta^2$, through the transformation
\[ \label{e:transform}
\psi=v_2-\ri v_1, \quad \phi=v_2+\ri v_1.
\]
This means that, in order for the type-I potential (\ref{e:V1}) to have all-real $\mu$ spectrum, the necessary and sufficient condition is that the $\zeta$ spectrum of the ZS spectral problem (\ref{ZS0}) is either real or purely imaginary (note that the continuous spectrum of the ZS problem is the real $\zeta$ axis). It was shown by Klaus and Shaw \cite{KlausShaw} that when $g(x)$ is a single-humped localized real function, then all discrete eigenvalues of the ZS problem are purely imaginary, and thus type-I potentials (\ref{e:V1}) have all-real spectra.

For type-II potentials (\ref{Eq:PotentialTypeII}), it was shown by Andrianov \emph{et al.} \cite{Andrianov} that if $c_2\le 0$, then the spectrum is strictly real. The proof is based on supersymmetry (see Sec. \ref{sec_SUSY}). Specifically, when $c_2=-\epsilon^2\le 0$, with $\epsilon$ being a real parameter, then we have the following intertwining operator relation,
\[ \label{e:intertwin}
\left[-\partial_x+W(x)\right]\left[\partial_{xx}+V(x)\right]=\left[\partial_{xx}+V_0(x)\right]\left[-\partial_x+W(x)\right],
\]
where $V(x)$ is the complex type-II potential (\ref{Eq:PotentialTypeII}),
\[
W(x)=\frac{g'+\epsilon}{2g}-\frac{1}{2}\ri g,   \nonumber
\]
and $V_0$ is a real potential,
\[
V_0(x)=\frac{1}{4}g^2+\frac{2gg''-3g'^2-4\epsilon g'-\epsilon^2}{4g^2}.  \nonumber
\]
The intertwining relation (\ref{e:intertwin}) shows that the Schr\"odinger operators $\partial_{xx}+V(x)$ and $\partial_{xx}+V_0(x)$ are related by a similarity transformation, and thus they share the same spectrum. Since the spectrum of the real potential $V_0$ is all-real, the spectrum of
the type-II potential (\ref{Eq:PotentialTypeII}) is then all-real as well. Note that for $c_2>0$, such an intertwining operator relation does not exist, and the supersymmetry approach does not apply. In such cases, phase transition can occur in type-II potentials as Fig. \ref{Fig:SecondOrder} shows.

\section{Non-\PT-symmetric potentials with all-real spectra and exceptional-point-free phase transition} \label{sec:type_new}

Extending the symmetry approach of the previous section, additional new types of complex potentials with all-real spectra can be constructed \cite{Yang2017}. More interestingly, these potentials exhibit exceptional-point-free phase transition, which is very novel.

In this construction, instead of choosing $\eta$ in Eq. (\ref{Eq:LSimilar}) as pure differential operators, we now take $\eta$ to be a combination of the parity operator $\P$ and differential operators. In the simplest case, we take $\eta$ to be a combination of the parity operator and a first-order differential operator, i.e.,
\begin{equation} \label{e:eta1b}
\eta =\P \left[\partial_x + h(x)\right],
\end{equation}
where $h(x)$ is a complex function to be determined. Substituting this $\eta$ into the
similarity condition (\ref{Eq:LSimilar}), we get the following two
equations
\begin{equation} \label{e:vpv}
V(x)-V^*(-x)=2h'(x),
\end{equation}
\begin{equation} \label{e:VV}
\left[V(x)-V^*(-x)\right]h(x)=h''(x)-V'(x).
\end{equation}
From the first equation, we see that $\left[h^*(-x)\right]_x=h'(x)$; thus
\begin{equation} \label{e:h}
h^*(-x)=h(x)+c_1,
\end{equation}
where $c_1$ is a constant. Substituting Eq. (\ref{e:vpv}) into (\ref{e:VV}) and integrating once, we get
\begin{equation} \label{e:Vx}
V(x)=h'(x)-h^2(x)+c_2,
\end{equation}
where $c_2$ is another constant. Lastly, inserting (\ref{e:h}) and (\ref{e:Vx}) into Eq. (\ref{e:vpv}), we obtain
\begin{equation} \label{e:c1c2}
c_1^2+2c_1h(x)+c_2-c_2^*=0.
\end{equation}
In order for the potential $V(x)$ in (\ref{e:Vx}) not to be a constant, the function $h(x)$ should not be identically zero. Thus, Eq. (\ref{e:c1c2}) dictates that $c_1=0$ and $c_2$ is real. The former condition means that the complex function $h(x)$ is \PT-symmetric in view of Eq. (\ref{e:h}). Regarding the latter condition, since a real constant in a potential can be easily removed by a simple shift of the eigenvalue, we can set $c_2=0$ without loss of generality. In the end, we find that for new complex potentials of the form
\begin{equation} \label{e:V}
V(x)=h'(x)-h^2(x),
\end{equation}
where $h(x)$ is a \PT-symmetric complex function, i.e., $h^*(x)=h(-x)$, the Schr\"odinger operator $L$ satisfies the similarity condition (\ref{Eq:LSimilar}) with $\eta$ given in (\ref{e:eta1b}). Because of this, the eigenvalues of $L$ exhibit complex-conjugate symmetry. Hence,
the spectrum of $L$ can be all-real, but phase transition may also occur, similar to \PT-symmetric potentials as well as non-\PT-symmetric potentials of the previous section.

In these new potentials, $h(x)$ is an arbitrary \PT-symmetric function. Because of this, simple algebra shows that these potentials can accommodate any arbitrary gain-loss profile \cite{Yang2017}, analogous to type-I and type-II potentials of the previous section.

A peculiar property of this new class of non-\PT-symmetric potentials is that, if these potentials are localized, then they will not admit any discrete real eigenvalues. This contrasts the previous non-\PT-symmetric potentials (\ref{e:V1}) and (\ref{Eq:PotentialTypeII}), where discrete real eigenvalues are very common (see Figs. 1 and 2).

To prove this statement, we recall that for any localized potential, the continuous spectrum of the Schr\"odinger operator $L$ is the semi-infinite interval $-\infty\le \mu\le 0$; and discrete real eigenvalues, if any, are positive numbers. Suppose $\mu=k^2$, with $k>0$, is a discrete real eigenvalue in the localized potential (\ref{e:V}). Since $L$ is a second-order differential operator, its discrete eigenvalue $\mu$ can only have geometric multiplicity one, meaning that the corresponding eigenfunction $\psi$ is unique (up to a constant multiple). Applying the operator $\eta$ to the eigenvalue equation $L\psi=k^2\psi$ and recalling the symmetry relation (\ref{Eq:LSimilar}), we get $L^*(\eta\psi)=k^2(\eta\psi)$. Taking the complex conjugate of this equation, we get $L(\eta\psi)^*=k^2(\eta\psi)^*$.
This means that $(\eta\psi)^*$ is also an eigenfunction of $L$ at the same eigenvalue $\mu$. Thus, $(\eta\psi)^*$ and $\psi$ must be linearly dependent on each other, i.e., $(\eta\psi)^*=\alpha\psi$,
where $\alpha$ is a complex constant. In view of the expression of $\eta$ in Eq.~(\ref{e:eta1b}), this relation can be rewritten as
\begin{equation} \label{psiconstraint}
[\partial_x +h(x)]\psi(x)=\alpha^*\psi^*(-x).
\end{equation}
Now we examine this relation as $x\to \pm \infty$. Since the potential $V(x)$ is localized, $h(x)$ is localized as well.
From the eigenvalue equation (\ref{Lpsi}), we see that the large-$x$ asymptotics of $\psi(x)$ is
\[
\psi(x) \to a_\pm e^{-k|x|}, \quad x\to \pm \infty,  \nonumber
\]
where $a_\pm$ are complex constants which cannot be both zero. Since $h(x)$ is localized, as $x\to \pm \infty$, the contribution of the $h(x)$ term in Eq.~(\ref{psiconstraint}) is subdominant and will be ignored. Then, inserting the above $\psi$-asymptotics into (\ref{psiconstraint}), we get two parameter conditions
\[
-ka_+=\alpha^*a_-^*, \quad ka_-=\alpha^*a_+^*.   \nonumber
\]
Dividing these two equations and rearranging terms, we get
\[
|a_+|^2+|a_-|^2=0,    \nonumber
\]
which is impossible since $a_\pm$ cannot be both zero. Thus, localized potentials (\ref{e:V}) do not admit discrete real eigenvalues.

The fact of localized potentials (\ref{e:V}) not admitting discrete real eigenvalues is a distinctive property, and it has important implications. Since there are no discrete real eigenvalues, a phase transition in these potentials clearly cannot be induced from collisions of such eigenvalues through an exceptional point. Instead, complex eigenvalues will have to bifurcate out from the continuous spectrum. Below, we will use an example to show that this is exactly the case. In this example, we take
\begin{equation} \label{e:hlocal}
h(x)=d_1 \hspace{0.02cm} \sech x+\ri \hspace{0.02cm} d_2 \hspace{0.03cm} \sech x \hspace{0.04cm} \tanh x,
\end{equation}
which is \PT-symmetric for real constants $d_1$ and $d_2$. We also fix $d_1=1$. Then for two different $d_2$ values of 1 and 2, the resulting non-\PT-symmetric localized potentials and their spectra are plotted in Fig. 3. We see that neither spectrum contains discrete real eigenvalues, which corroborates our analytical result above. When $d_2=1$, the spectrum is all-real (see the upper right panel). But when $d_2=2$, a conjugate pair of discrete eigenvalues $\mu\approx -0.7067\pm 0.4961\ri$ appear (see the lower right panel). The phase transition occurs at $d_2\approx 1.385$. Closer examination reveals that the two complex eigenvalues bifurcate out from $\mu_0\approx -0.8062$, which is in the interior of the continuous spectrum. It is also noticed that the discrete (localized) eigenfunctions of the two complex eigenvalues bifurcate out from two different continuous (nonlocal) eigenfunctions of the real eigenvalue $\mu_0$, rather than from a single coalesced eigenfunction. This reveals two facts: (1) these discrete eigenmodes bifurcate out from continuous eigenmodes, rather than embedded isolated eigenmodes, in the interior of the continuous spectrum; (2) this phase transition does not go through an exceptional point. The second fact is particularly significant, because all phase transitions reported before in both finite- and infinite-dimensional non-Hermitian systems occurred either due to a collision of real eigenvalues forming an exceptional point, where different eigenvectors or eigenfunctions coalesce \cite{Bender1998,Ahmed2001,Yang2016}, or through an exotic singular scenario, where complex eigenvalues bifurcate out from infinity of the real axis \cite{Konotopsingular}. This is the first instance where a phase transition occurs without an exceptional point or a singular point. Very recently, an analytical explanation of this mysterious bifurcation was given by Konotop and Zezyulin \cite{Konotop2017} through the splitting of self-dual spectral singularity.

\begin{figure}[tb!]
\begin{center}
\includegraphics[width=0.8\textwidth]{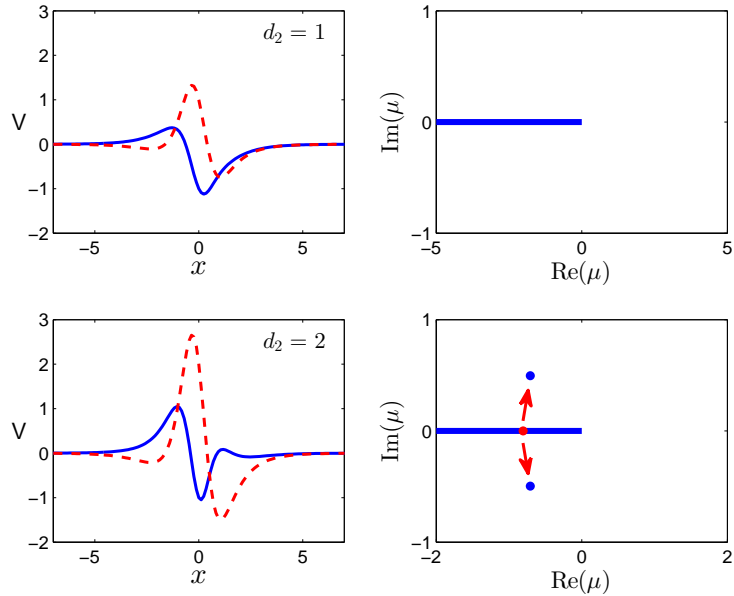}
\end{center}
\smallskip
\caption{Spectra of localized potentials (\ref{e:V}) with $h(x)$ given in (\ref{e:hlocal}) and $d_1=1$ (the $d_2$ values are shown inside the panels). Left column: real (solid blue) and imaginary (dashed red) parts of the complex potentials. Right column: spectra of potentials in the left column (the red arrows in the lower panel indicate that the two complex eigenvalues in the spectrum bifurcate out from the red dot in the interior of the continuous spectrum when a phase transition happens). Adapted from \cite{Yang2017}.}
\label{Fig3}
\end{figure}

\section{Construction of non-\PT-symmetric potentials with all-real spectra using supersymmetry} \label{sec_SUSY}

The concept of supersymmetry (SUSY) was first introduced in quantum field theories and high-energy physics (see \cite{Cooper1995} and
the references therein). Subsequently, SUSY was utilized in quantum mechanics to construct analytically solvable potentials. This
construction is based on the factorization of the Schr\"odinger operator into the product of two first-order
operators. Switching the order of these two first-order operators gives another Schr\"odinger operator with a new potential (called the {\em partner potential}) which shares the same spectrum as the original potential (except possibly a single
discrete eigenvalue). SUSY can establish perfect phase matching between modes in the partner potentials, which has motivated applications such as mode converters in SUSY optical structures \cite{SUSY_experiment}. Extending the idea of SUSY, parametric families of complex potentials with all-real spectra can be constructed~\cite{Andrianov,Khare89,Cannata98,Bagchi01,Miri2013}.

Let us employ the idea of SUSY to construct complex potentials with all-real spectra, following \cite{Khare89,Miri2013,Yang2014}.

Suppose $V_1(x)$ is a potential with all-real spectrum, and
$\mu^{(1)}$ is an eigenvalue of this potential with eigenfunction
$\psi^{(1)}$, i.e.,
\[ \label{e:eigV1}
\left[ \frac{d^2}{dx^2} +V_1(x)-\mu^{(1)}\right] \psi^{(1)}=0.
\]
We first factorize the linear operator in this equation as
\[   \label{e:factorV1}
-\frac{d^2}{dx^2} -V_1(x)+\mu^{(1)} = \left[-\frac{d}{dx}+W(x)\right]\left[\frac{d}{dx}+W(x)\right].
\]
The function $W(x)$ in this factorization can be obtained by
requiring $\psi^{(1)}$ to annihilate $d/dx+W(x)$, and this gives
$W(x)$ as
\[  \label{f:W}
W(x)=-\frac{d}{dx} \ln (\psi^{(1)}).
\]
It is easy to directly verify that this $W(x)$ does satisfy the
factorization equation~(\ref{e:factorV1}).

Now we switch the two operators on the right side of the above
factorization, and this leads to a new potential $V_2(x)$,
\[  \label{e:factorV2}
-\frac{d^2}{dx^2} -V_2(x)+\mu^{(1)} = \left[\frac{d}{dx}+W(x)\right]\left[-\frac{d}{dx}+W(x)\right],
\]
where
\[  \label{f:V2}
V_2=V_1-2W_x.
\]
This $V_2$ potential is referred to as the partner potential of $V_1$. It is known that for any two operators $A$ and $B$,  $AB$ and $BA$ share the same spectrum in general (except for a possible difference in the zero eigenvalue when the kernel of $A$ or $B$ is non-empty). Then, in view of the two factorizations (\ref{e:factorV1}) and (\ref{e:factorV2}), we see that the spectrum of $V_2$ is that of $V_1$, but with $\mu^{(1)}$ generically removed.

The new potential $V_2$, however, is only real or \PT-symmetric if
$V_1$ is so. In order to derive non-\PT-symmetric potentials, we
build a new factorization for the $V_2$ potential,
\[  \label{e:factorV2new}
-\frac{d^2}{dx^2}-V_2(x)+\mu^{(1)} =
\left[\frac{d}{dx}+\widetilde{W}(x)\right]\left[-\frac{d}{dx}+\widetilde{W}(x)\right].
\]
The function $\widetilde{W}$ in this new factorization can be derived as follows
\cite{Khare89,Miri2013}. Equating this new factorization with the previous one in (\ref{e:factorV2}), we get
\[
\widetilde{W}_x+\widetilde{W}^2=W_x+W^2.   \nonumber
\]
Denoting $\widetilde{W}=W+\phi$, we see $\phi$ satisfies a Ricatti equation
\[
\phi_x+2W\phi+\phi^2=0.   \nonumber
\]
Through the standard variable transformation $\phi=q'/q$, the function $q$ is found to satisfy a linear homogeneous equation
\[
q''+2Wq'=0.   \nonumber
\]
Utilizing the $W$ expression in (\ref{f:W}), we obtain the general $q$ solution as
\[
q=\hat{c}\left[c+\int_0^x [\psi^{(1)}(\xi)]^2 d\xi\right],   \nonumber
\]
where $c$ and $\hat{c}$ are arbitrary complex constants. In view of the variable transformation $\phi=q'/q$, we see the constant $\hat{c}$ does not contribute to the $\phi$ solution. Putting all the above results together, we find the general function $\widetilde{W}(x)$ as
\[  \label{f:Wtilde}
\widetilde{W}(x)=-\frac{d}{dx} \ln (\widetilde{\psi}^{(1)}),
\]
where
\[
\widetilde{\psi}^{(1)}(x) = \frac{\psi^{(1)}(x)}{c+\int_0^x [\psi^{(1)}(\xi)]^2 d\xi}.   \nonumber
\]
For the new $V_2$ factorization (\ref{e:factorV2new}), its partner potential, defined through
\[
-\frac{d^2}{dx^2} -\widetilde{V}_1(x)+\mu^{(1)} = \left[-\frac{d}{dx}+\widetilde{W}(x)\right]\left[\frac{d}{dx}+\widetilde{W}(x)\right],  \nonumber
\]
is
\[
\widetilde{V}_1=V_2+2\widetilde{W}_x.   \nonumber
\]
Utilizing the $V_2$ and $\widetilde{W}$ formulae (\ref{f:V2}) and
(\ref{f:Wtilde}), this $\widetilde{V}_1$ potential is then found to
be
\[  \label{f:V1tilde}
\widetilde{V}_1(x)=V_1(x)+2\frac{d^2}{dx^2}\ln\left[c+\int_0^x
[\psi^{(1)}(\xi)]^2 d\xi\right].
\]
For generic values of the complex constant $c$, this
$\widetilde{V}_1$ potential is complex and non-\PT-symmetric. In
addition, its spectrum is identical to that of $V_1$. Indeed, even
though $\mu^{(1)}$ may not lie in the spectrum of $V_2$, it is in
the spectrum of $\widetilde{V}_1$ with eigenfunction
$\widetilde{\psi}^{(1)}$. Hence, if $V_1$ has an all-real spectrum,
so does $\widetilde{V}_1$. Notice that this $\widetilde{V}_1$
potential, referred to as the superpotential below, is actually a family
of potentials due to the free complex constant $c$.

\begin{figure}
    \includegraphics[width=\columnwidth]{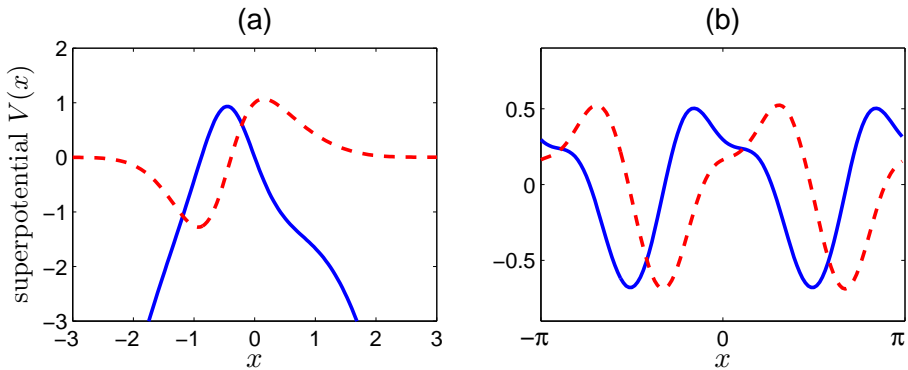}
    \caption{(a) Superpotential (\ref{f:Vexample1}) with $c=1+\ri$;
        (b) Periodic superpotential (\ref{f:Vexample3}) with $V_0=1$ and $c=0.5-2\ri$. The solid blue curve is Re($V$), and the dashed red curve is Im($V$). Adapted from \cite{Yang2014}. }  \label{fig_SUSY}
\end{figure}

Now we give two explicit examples of non-\PT-symmetric
superpotentials (\ref{f:V1tilde}) with all-real spectra. The first
one is constructed from the parabolic potential $V(x)=-x^2$ and its
first eigenmode of $\mu_1=-1$ with $\psi_1=e^{-x^2/2}$. Then the
superpotential (\ref{f:V1tilde}) reads
\begin{equation}  \label{f:Vexample1}
V(x)=-x^2+2\frac{d^2}{dx^2}\ln\left[c+\int_0^x e^{-\xi^2}d\xi\right].
\end{equation}
This potential with $c=1+\ri$ is shown in Fig.~\ref{fig_SUSY}(a). The spectrum of this potential
(for any complex $c$ value) is $\{-1, -3, -5, \dots\}$, i.e.,  is all-real.

In the second example, the superpotential (\ref{f:V1tilde}) is built
from the \PT-symmetric periodic potential $V(x)=V_0^2 e^{2\ri x}$
and its Bloch mode $\psi^{(1)}=I_1(V_0e^{\ri x})$ with eigenvalue
$\mu_1=-1$. Here $V_0$ is a real constant, and $I_1$ is the
modified Bessel function. The resulting periodic superpotential
(\ref{f:V1tilde}) reads
\begin{equation}  \label{f:Vexample3}
V(x)=V_0^2 e^{2\ri x}+2\frac{d^2}{dx^2}\ln\left[c+\int_0^x  I_1^2(V_0
e^{\ri\xi}) d\xi\right].
\end{equation}
For $V_0=1$ and $c=0.5-2\ri$, this potential is shown in Fig.~\ref{fig_SUSY}(b).
The diffraction (dispersion) relation of this
superpotential (for all $c$ values) is the same as that of the
original potential $V(x)=V_0^2 e^{2\ri x}$, i.e., $\mu=-(k+2m)^2$,
where the wavenumber $k$ is in the first Brillouin zone, $k\in [-1, 1]$, and $m$ is any
non-negative integer.

If $V(x)$ is a localized real potential, then SUSY allows to construct localized
complex superpotentials (\ref{f:V1tilde}) with all-real spectra  \cite{Miri2013,Yang2014}.

A related technique to construct complex potentials with all-real spectra was proposed by Cannata, \emph{et al.} \cite{Cannata98}. This technique is based on the formulae (\ref{f:W}) and (\ref{f:V2}). But instead of choosing $\mu^{(1)}$ as an eigenvalue of the potential $V_1(x)$ and $\psi^{(1)}$ as the corresponding eigenfunction, one chooses $\mu^{(1)}$ as an arbitrary real number and
the function $\psi^{(1)}$ as a complex linear combination of the two fundamental solutions to the Schr\"odinger equation (\ref{e:eigV1}) [here we do not require $\psi^{(1)}$ to be square-integrable]. For instance, if $V_1(x)$ is a real potential and $\mu^{(1)}$ is an arbitrary real number, then we can choose $\psi^{(1)}$ as a linear combination $c_1f_1(x)+c_2f_2(x)$, where $f_1, f_2$ are the two real fundamental solutions to Eq. (\ref{e:eigV1}), and $c_1, c_2$ are complex constants. With such choices of $\mu^{(1)}$ and $\psi^{(1)}$, it is easy to see that the potential $V_1(x)$ and the complex potential $V_2(x)$ [as given by Eq. (\ref{f:V2})] still share the same spectrum in general. The only possible exception is regarding $\mu^{(1)}$. If $1/\psi^{(1)}$ is square-integrable, then since
\[
\left[-\frac{d}{dx}+W(x)\right]\frac{1}{\psi^{(1)}}=0,  \nonumber
\]
$\mu^{(1)}$ is in the discrete spectrum of $V_2$; but it may not be in the discrete spectrum of $V_1$. Using this construction, non-\PT-symmetric complex potentials with all-real spectra can also be obtained. For examples, see \cite{Cannata98}.

One more variation of SUSY is based on the following observation. It can be seen from Eqs. (\ref{e:factorV1}) and (\ref{e:factorV2}) that, for any complex functions $W(x)$ and a complex constant $c$, the two potentials
\[
-V_1(x)=W^2(x)-W'(x)+c, \quad -V_2(x)=W^2(x)+W'(x)+c,  \nonumber
\]
form partner potentials which share the same spectrum (with the possible exception of a single bound state). Thus, if we choose $W(x)$ so that $V_1(x)$ is real, then the resulting complex potential $V_2(x)$ will have an all-real spectrum. These complex potentials $V_2(x)$ turn out to be type-II potentials (\ref{Eq:PotentialTypeII}) described in Sec. \ref{sec_typeII} but with $c_2\le 0$. An equivalent derivation of this result was given by Andrianov \emph{et al.} \cite{Andrianov} and reviewed in the end of Sec. \ref{sec_typeI_II_all_real}.

\section{Construction of non-\PT-symmetric potentials with all-real spectra using soliton theory} \label{sec:soliton}
Another technique to construct complex potentials with all-real spectra is to use the soliton theory. This technique was proposed by Wadati \cite{Wadati2008} for the construction of \PT-symmetric potentials with all-real spectra, but it apparently can be extended to construct non-\PT-symmetric potentials with all-real spectra, as we will demonstrate below.

Let us consider the modified Korteweg-de Vries (mKdV) equation
\begin{eqnarray}
\label{mKdV}
u_t+6u^2u_{x}+u_{xxx}=0
\end{eqnarray}
for the real function $u(x,t)$, where $x$ is the spatial coordinate, and $t$ is time. We will consider localized solutions, $\lim_{|x|\to\infty}u(x,t)=0$. Equation (\ref{mKdV}) is the
compatibility condition between the Zakharov-Shabat (ZS) spectral problem  \cite{ZS71}
\begin{equation}
\label{ZS}
v_{1x}+\ri\zeta v_1=u(x,t)v_2, \,\,\, v_{2x}-\ri\zeta v_2=-u(x,t)v_1,
\end{equation}
and the linear system
\begin{eqnarray*}
    v_{1t}=2\ri\zeta(u^2-2\zeta^2)v_1+(2\ri\zeta u_x-2u^3-u_{xx}+4\zeta^2 u)v_2,
    \\
    v_{2t}= (2\ri\zeta u_x+2u^3+u_{xx}-4\zeta^2 u)v_1-2\ri\zeta(u^2-2\zeta^2)v_2.
\end{eqnarray*}
Here, $\zeta$ is a spectral parameter.

The ZS spectral problem (\ref{ZS}) can be transformed into Schr\"odinger eigenvalue problems through the transformation (\ref{e:transform}). Under this transformation, we get
\[
\psi_{xx}+V(x,t)\psi=\mu\psi,  \quad  \phi_{xx}+V^*(x,t)\phi=\mu \phi,
\]
where
\[ \label{VW}
V(x,t) =u^2(x,t)+iu_x(x,t),
\]
and $\mu=-\zeta^2$. Here, time $t$ plays the role of a parameter. If $u(x,t)$ is an even function of $x$, then the potential $V(x,t)$ is \PT symmetric; for general $u(x,t)$ solutions, this potential is complex and non-\PT-symmetric.

Discrete eigenvalues of the ZS problem (\ref{ZS}) appear as quadruples $(\zeta, \zeta^*, -\zeta, -\zeta^*)$ if $\zeta$ is complex and as complex-conjugate pairs $(\zeta, \zeta^*)$ if $\zeta$ is purely imaginary. The continuous spectrum of the ZS problem is the real-$\zeta$ axis. In view of the above connection between the ZS and Schr\"odinger eigenvalue problems, we see that from {\em any} solution $u(x,t)$ of the mKdV equation (\ref{mKdV}) that possesses purely imaginary discrete ZS eigenvalues, one can obtain a complex potential $V(x,t)$, defined by (\ref{VW}), with strictly real spectrum. Further, we notice that while $u(x,t)$ depends on the parameter $t$,
its ZS spectrum does not since $u(x,t)$ satisfies the mKdV equation. This means that $t$ can be considered as a ``deformation'' parameter, and $u(x,t)$  generates a family of deformable potentials $V(x,t)$ with the same real spectrum. Since the solution $u(x,t)$ is asymmetric in general, the resulting complex potential $V(x,t)$ is then non-\PT-symmetric.

Analytical solutions $u(x,t)$ with purely imaginary discrete ZS eigenvalues can be derived by the soliton theory. Indeed, through the inverse scattering method, $N$-solitons of the mKdV equation with purely imaginary discrete eigenvalues $\{\pm \zeta_n, 1\le n\le N\}$ were found as \cite{Wadati_soliton}
\[
u(x,t)=-2\frac{\partial}{\partial x}\mbox{arctan}\frac{\mbox{Im}\mbox{det}(I+A)}{\mbox{Re}\mbox{det}(I+A)},
\]
where $I$ is the $N\times N$ identity matrix, Re and Im represent the real and imaginary parts, $A$ is the $N\times N$ matrix whose elements are
\[
A_{nm}(x,t)=-\frac{c_n}{\zeta_n+\zeta_m}e^{\ri(\zeta_n+\zeta_m)x+8\ri\zeta_n^3t},   \nonumber
\]
$\zeta_n=\ri\eta_n$, $\eta_n>0$, and $c_n$ are real constants. The corresponding complex potential $V(x,t)$ from Eq. (\ref{VW}) then would have all-real spectrum, with discrete eigenvalues as $-\zeta_n^2=\eta_n^2 \  (1\le n\le N)$ and the continuous spectrum as $(-\infty, 0]$.

\begin{figure}
    \includegraphics[width=\columnwidth]{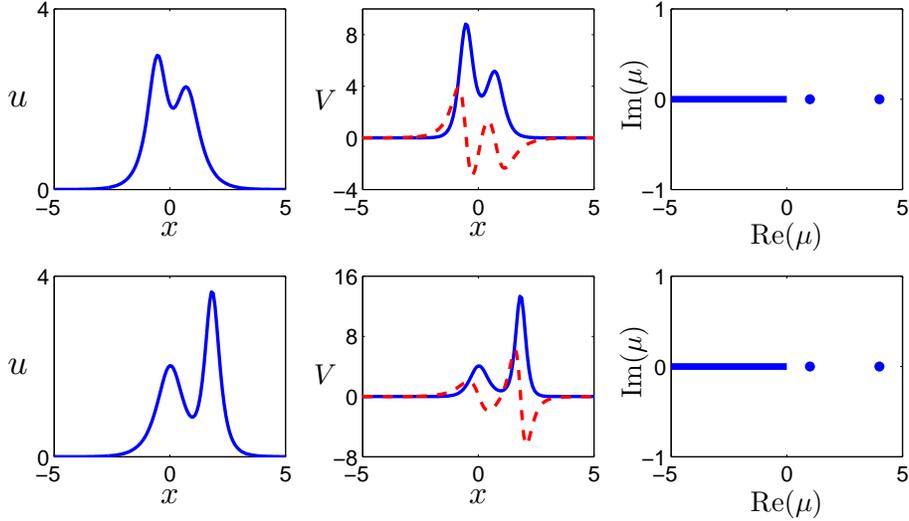}
    \caption{Non-\PT-symmetric potentials with real spectra from the soliton theory. Left column: the two-soliton solution (\ref{u_Wadati}) with parameters (\ref{para_Wadati}); middle column: the corresponding complex potential $V(x,t)$ from Eq. (\ref{VW}); right column: spectrum of this $V(x,t)$ potential. Upper row: $t=0$; lower row: $t=0.1$.  }  \label{fig_Wadati}
\end{figure}

As an example, we present non-\PT-symmetric potentials obtained from the two-soliton solution of the mKdV equation. These two solitons are found from the above general formula by taking $N=2$ and can be written as \cite{Wadati_soliton}
\[ \label{u_Wadati}
u(x, t)=4\frac{\eta_1+\eta_2}{\eta_2-\eta_1}\, \frac{G(x,t)}{F(x,t)},
\]
where
\[G(x,t)=
\epsilon_1\eta_1\cosh\left[2\eta_2x +\delta_2(t)+\frac{1}{2}\gamma_{12}\right]+
\epsilon_2\eta_2\cosh\left[2\eta_1x +\delta_1(t)-\frac{1}{2}\gamma_{12}\right],   \nonumber
\]
\begin{eqnarray*}
F(x,t)=& \hspace{-1cm}   \cosh[2(\eta_1+\eta_2)x+\delta_1(t)+\delta_2(t)]+\frac{4\eta_1\eta_2\epsilon_1\epsilon_2}{(\eta_1-\eta_2)^2}  \\
& +\left(\frac{\eta_1+\eta_2}{\eta_1-\eta_2}\right)^2\cosh[2(\eta_2-\eta_1)x+\delta_2(t)-\delta_1(t)+\gamma_{12}],
\end{eqnarray*}
\[
\epsilon_1=\pm 1, \quad \epsilon_2=\pm 1, \quad \eta_1>0, \quad \eta_2>0,  \nonumber
\]
\[
\delta_1(t)=\delta_1-8\eta_1^3t, \quad \delta_2(t)=\delta_2-8\eta_2^3t, \quad \gamma_{12}=\ln(\eta_2/\eta_1),   \nonumber
\]
and $\delta_1, \delta_2$ are real constants. To illustrate, we take
\[ \label{para_Wadati}
\eta_1=1,  \quad \eta_2=2, \quad  \delta_1=\delta_2=0, \quad \epsilon_1=\epsilon_2=1.
\]
The soliton $u(x,t)$, the complex potential $V(x,t)$ and its spectrum at times $t=0$ and 0.1 are displayed in the upper and lower rows of Fig. \ref{fig_Wadati} respectively. Both complex potentials are non-\PT-symmetric and differ from each other significantly, but they have identical real spectra.

\section{Partially-\PT-symmetric potentials in multi-dimensions}

In this section, we consider the generalization of \PT symmetry to higher spatial dimensions. Let us consider a (2+1)-dimensional generalization of the paraxial linear beam propagation equation (\ref{e:SE}),
\begin{equation} \label{Eq:NLS}
\ri \Psi_z + \Psi_{xx}+\Psi_{yy} + V(x,y)\Psi = 0,
\end{equation}
where $z$ is the propagation direction, and $(x,y)$ is the transverse
plane. Looking for eigenmodes of the form $\Psi(x,y,z) = \re^{\ri \mu z} \psi(x,y)$ we arrive at the eigenvalue problem
\begin{equation} \label{e:eigen}
(\partial_{xx}+\partial_{yy}+V)\psi=\mu\psi,
\end{equation}
where $\mu$ is the eigenvalue and $\psi$ the eigenfunction.

The usual \PT symmetry of the complex potential $V(x,y)$ is defined as
\[
V^*(x,y)=V(-x,-y),
\]
i.e., the potential is invariant under complex conjugation and simultaneous
reflections in both $x$ and $y$ directions. For these potentials, the spectrum can be all-real, with a possibility of phase transition, just like one-dimensional \PT-symmetric potentials.

However, this usual concept of \PT symmetry can be generalized. Indeed, if the potential is invariant under complex conjugation and
reflection in a \emph{single} spatial direction, i.e.,
\begin{equation} \label{e:PPTcondition}
V^*(x,y)=V(-x,y), \hspace{0.15cm} \mbox{or}\hspace{0.28cm} V^*(x,y)=V(x,-y),
\end{equation}
its spectrum can still be all-real with a possibility of phase transition. These potentials were introduced in \cite{YangPPT} and termed partially-\PT-symmetric potentials.

The fundamental reason these partially-\PT-symmetric potentials can also feature all-real spectra is that, their eigenvalues also come in complex conjugate pairs ($\mu, \mu^*$). This eigenvalue symmetry is a common feature of \PT-symmetric potentials, partially-\PT-symmetric potentials, and complex potentials derived in sections \ref{sec:typeI_II} and \ref{sec:type_new}, which results in the possibility of all-real spectra for all these potentials.

The complex-conjugate-pair eigenvalue symmetry for these partially-\PT-symmetric potentials is easy to prove. Indeed, if $V^*(x,y)=V(-x,y)$ or $V^*(x,y)=V(x,-y)$, then by taking the complex conjugate of Eq. (\ref{e:eigen}) and switching $x\to -x$ or $y\to -y$, we see that $\mu^*$ would also be an eigenvalue with eigenfunction $\psi^*(-x,y)$ or $\psi^*(x,-y)$.

\begin{figure}
\begin{center}
\includegraphics[width=0.8\textwidth]{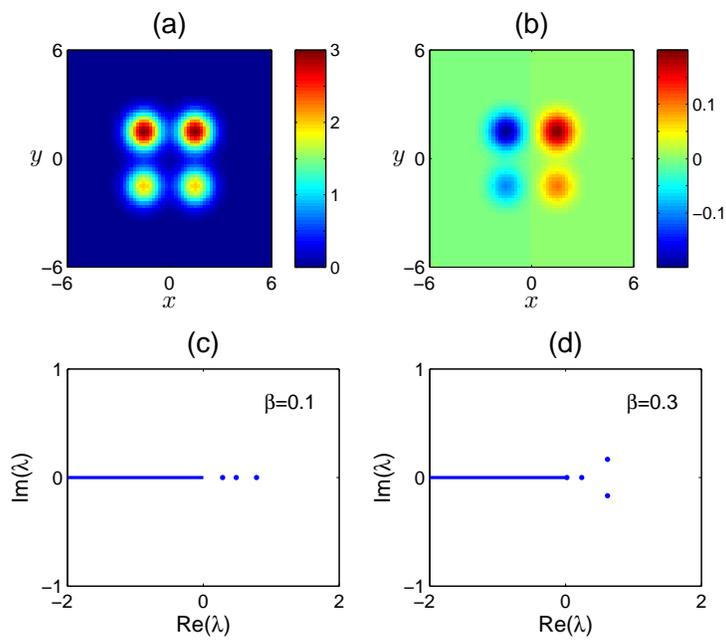}
\caption{(a,b) Real and imaginary parts of the partially-\PT-symmetric potential
(\ref{e:Vexample}) for $\beta=0.1$; (c, d) spectrum of this potential for $\beta=0.1$ and 0.3 respectively. Adapted from \cite{YangPPT}. } \label{f:PPT}
\end{center}
\end{figure}

As an example, we take the partially-\PT-symmetric complex potential $V(x,y)$ to be localized at four spots:
\begin{eqnarray} \label{e:Vexample}
V(x,y)=3\left(e^{-(x-x_0)^2-(y-y_0)^2}+e^{-(x+x_0)^2-(y-y_0)^2}\right) \nonumber \\ +2\left(e^{-(x-x_0)^2-(y+y_0)^2}+e^{-(x+x_0)^2-(y+y_0)^2}\right)   \nonumber \\
+\ri\beta \left[2\left(e^{-(x-x_0)^2-(y-y_0)^2}-e^{-(x+x_0)^2-(y-y_0)^2}\right) \right. \nonumber \\ \left.  +\left(e^{-(x-x_0)^2-(y+y_0)^2}-e^{-(x+x_0)^2-(y+y_0)^2}\right)\right],
\end{eqnarray}
where $x_0, y_0$ control the separation distances between these four
spots, and $\beta$ is a real constant. For definiteness, we set
$x_0=y_0=1.5$. This potential is not \PT-symmetric, but is
partially-\PT-symmetric with symmetry $V^*(x,y)=V(-x,y)$. For
$\beta=0.1$, this potential is displayed in Fig. \ref{f:PPT} (top row). It is
seen that Re($V$) is symmetric in $x$, Im($V$) anti-symmetric in
$x$, and both Re($V$), Im($V$) are asymmetric in $y$. The spectrum
of this potential is plotted in Fig. \ref{f:PPT}(c). It is seen that this
spectrum contains three discrete eigenvalues and the continuous
spectrum, which are all-real.

For potential (\ref{e:Vexample}) with varying $\beta$, we have found
that its spectrum is all-real as long as $|\beta|$ is below a
threshold value of $0.214$. Above this threshold, a phase transition
occurs, where pairs of real eigenvalues coalesce and then bifurcate off into the complex plane. This phase
transition is illustrated in Fig. \ref{f:PPT}(d), where the spectrum at
$\beta=0.3$ is shown.

\section{Summary and discussion}

In this article, we have reviewed various approaches to generalize \PT symmetry. We have shown that large classes of non-\PT-symmetric complex potentials can also feature all-real spectra. These potentials are constructed by a variety of techniques, such as the symmetry method, the supersymmetry method, the soliton theory and partial \PT symmetry. Of these non-\PT-symmetric potentials, the ones derived from the symmetry condition (\ref{Eq:LSimilar}) in sections \ref{sec:typeI_II} and \ref{sec:type_new} allow for arbitrary gain-loss profiles. In addition, as free parameters and functions in those potentials vary, the spectrum could change, and phase transition (either through exceptional points or without) can occur. In non-\PT-symmetric potentials derived from supersymmetry and the soliton theory, on the other hand, the gain-loss profile is not totally free; and as free parameters in those potentials vary, the spectrum stays exactly the same.

The focus of this article was the spectrum of non-\PT-symmetric complex potentials, which is inherently a linear theory. When the spectrum of the complex potential is all-real, then wave propagation in the linear evolution equations (\ref{e:SE}) and (\ref{Eq:NLS}) would show features which resemble those in real potentials (without gain and loss). When nonlinearity arises in these complex potentials, where nonlinear terms appear in the evolution equations (\ref{e:SE}) and (\ref{Eq:NLS}), the interplay between nonlinearity and these complex potentials is an interesting question. In \PT-symmetric potentials and other \PT-symmetric systems, this interplay between nonlinearity and \PT symmetry has been reviewed in \cite{Yang_review,Kivshar_review}. In non-\PT-symmetric potentials, it was shown that the evolution equation (\ref{e:SE}) with Kerr nonlinearity could admit continuous families of solitons in type-I potentials (\ref{e:V1}), but not in
other types of complex potentials \cite{Tsoy2014,Yang2014,Konotop2014,YangSAPM2016}. How other types of nonlinearities interact with these complex potentials is a worthy question for study in the future.

\section*{Acknowledgment}
This material is based upon work supported by the Air Force Office of Scientific Research under award number FA9550-18-1-0098, and the National Science Foundation under award number DMS-1616122.


\begin{thebibliography}{10}

\bibitem{Kivshar_book} Y.S. Kivshar and G.P. Agrawal, \emph{Optical Solitons: From Fibers to Photonic Crystals}
(Academic Press, San Diego, 2003).

\bibitem{Musslimani2008}
Z.H. Musslimani, K.G. Makris, R. El-Ganainy and D.N.
Christodoulides,
``Optical solitons in \PT periodic potentials",
{Phys. Rev. Lett.} 100, 030402 (2008).

\bibitem{Yang_book} J. Yang, \emph{Nonlinear Waves in Integrable and Nonintegrable Systems} (SIAM, Philadelphia, 2010).

\bibitem{Bender1998} C.M. Bender and S. Boettcher,
``Real spectra in non-Hermitian Hamiltonians having \PT symmetry",
{Phys. Rev. Lett.} 80, 5243--5246 (1998).

\bibitem{Ali} A. Mostafazadeh,
``Pseudo-Hermitian representation of quantum mechanics",
{Int. J. Geom. Meth. Mod. Phys.} 7, 1191--1306 (2010).

\bibitem{BECbook} L.P. Pitaevskii and S. Stringari, \emph{Bose-–Einstein
Condensation} (Oxford University Press, Oxford, 2003).

\bibitem{Ahmed2001} Z. Ahmed,
``Real and complex discrete eigenvalues in an exactly solvable one-dimensional complex \PT-invariant potential",
Phys. Lett. A 282, 343--348 (2001).

\bibitem{Feng2013}
L. Feng, Y.L. Xu,  W.S. Fegadolli, M.H. Lu, J.E.B. Oliveira, V.R.
Almeida, Y.F. Chen, and A. Scherer,
``Experimental demonstration of a unidirectional reflectionless parity-time metamaterial at optical frequencies",
Nature Materials, 12, 108--113 (2013).

\bibitem{PTlaser_Zhang} L. Feng, Z.J. Wong, R. Ma, Y. Wang, and X. Zhang,
``Single-mode laser by parity-time symmetry breaking",
Science 346, 972--975 (2014).

\bibitem{PTlaser_CREOL} H. Hodaei, M.-A. Miri, M. Heinrich, D. N. Christodoulides, and M.
Khajavikhan,
``$\mathcal{PT}$-symmetric micro-ring laser",
Science 346, 975--978 (2014).

\bibitem{Peng}
B. Peng, S.K. \"Ozdemir, F. Lei, F. Monifi, M. Gianfreda, G.L. Long,
S. Fan, F. Nori, C.M. Bender, and L. Yang, ``Parity–time-symmetric whispering-gallery microcavities", Nat. Phys. 10,
394--398 (2014).

\bibitem{Yang_review}
V.V. Konotop, J. Yang and D.A. Zezyulin, ``Nonlinear waves in \PT-symmetric systems", Rev. Mod. Phys. 88, 035002 (2016).

\bibitem{Yang2016}
S. Nixon and J. Yang, ``All-real spectra in optical systems with arbitrary gain and loss distributions," Phys. Rev. A 93, 031802(R) (2016).

\bibitem{Tsoy2014} {E.N. Tsoy, I.M. Allayarov and F. Kh. Abdullaev,}
``Stable localized modes in asymmetric waveguides with gain and loss",
{Opt. Lett.} 39, 4215--4218 (2014).

\bibitem{Wadati2008} M. Wadati,
``Construction of parity-time symmetric potential through the soliton theory",
J. Phys. Soc. Jpn. 77, 074005 (2008).

\bibitem{Andrianov}
A.A. Andrianov, M.V. Ioffe, F. Cannata, and J.P. Dedonder,
``SUSY quantum mechanics with complex superpotentials and real energy spectra",
Int. J. Mod. Phys. A 14, 2675--2688 (1999).

\bibitem{ZS71}
V.E. Zakharov and A.B. Shabat, ``Exact theory of two-dimensional self-focusing and
one-dimensional self-modulation of waves in nonlinear media," Zh. E'ksp. Teor. Fiz. 61,
118 (1971) [Sov. Phys. JETP 34, 62--69 (1972)].

\bibitem{KlausShaw}
M. Klaus and J.K. Shaw, ``Purely imaginary eigenvalues of Zakharov-Shabat systems,"
Phys. Rev. E 65, 036607 (2002).

\bibitem{Yang2017}
J. Yang, ``Classes of non-parity-time-symmetric optical potentials with exceptional-point-free phase transitions", Opt. Lett. 42, 4067--4070 (2017).

\bibitem{Konotopsingular}
Y.V. Kartashov, V.V. Konotop, and D.A. Zezyulin,
``$\mathcal{CPT}$-symmetric spin-orbit–coupled condensate",
Europhys. Lett. 107, 50002 (2014).

\bibitem{Konotop2017}
V.V. Konotop and D.A. Zezyulin, ``Phase transition through the
splitting of self-dual spectral singularity in optical potentials," Opt. Lett. 42,
5206--5209 (2017).

\bibitem{Cooper1995}
F. Cooper, A. Khare and U. Sukhatme,
``Supersymmetry and quantum mechanics",
Phys. Rep. 251, 267--385 (1995).

\bibitem{SUSY_experiment}
M. Heinrich, M.-A. Miri, S. St\"utzer, R. El-Ganainy, S. Nolte, A. Szameit, and D. N. Christodoulides,
``Supersymmetric mode converters," Nature Commun. 5, 3698 (2014).

\bibitem{Khare89}
A. Khare and U. Sukhatme, ``Phase-equivalent potentials obtained from
supersymmetry", J. Phys. A 22, 2847--2860 (1989).

\bibitem{Cannata98}
F. Cannata, G. Junker and J. Trost,
``Schr\"odinger operators with complex potential but real spectrum",
Phys. Lett. A 246, 219--226 (1998).

\bibitem{Bagchi01}
B. Bagchi, S. Mallik, and C. Quesne,
``Generating complex potentials with real eigenvalues in supersymmetric quantum mechanics",
Int. J. Mod. Phys. A 16, 2859 (2001).

\bibitem{Miri2013}
M. Miri, M. Heinrich and D.N. Christodoulides,
``Supersymmetry-generated complex optical potentials with real spectra",
Phys. Rev. A 87, 043819 (2013).

\bibitem{Yang2014}
J. Yang, ``Necessity of \PT symmetry for soliton families in one-dimensional complex potentials", Phys. Lett. A 378, 367--373 (2014).

\bibitem{Wadati_soliton}
M. Wadati and K. Ohkuma, ``Multiple-pole solutions of the modified Korteweg-de Vries equation",
J. Phys. Soc. Jpn. 51, 2029--2035 (1982).

\bibitem{YangPPT}
J. Yang, ``Partially \PT-symmetric optical potentials with all-real spectra and soliton families in multi-dimensions", Opt. Lett. 39, 1133--1136 (2014).

\bibitem{Kivshar_review}
S.V. Suchkov, A.A. Sukhorukov, J. Huang, S.V. Dmitriev, C. Lee, and Yu. S. Kivshar,
``Nonlinear switching and solitons in \PT-symmetric photonic systems",
Laser Photonics Rev. 10, 177--213 (2016).

\bibitem{Konotop2014}
V.V. Konotop and D.A. Zezyulin, ``Families of stationary modes in complex potentials", Opt. Lett. 39, 5535--5538 (2014).

\bibitem{YangSAPM2016}
S. Nixon and J. Yang, ``Bifurcation of soliton families from linear modes in non-\PT-symmetric complex potentials", Stud. Appl. Math. 136, 459--483 (2016).




\end{thebibliography}
\end{document}